\documentclass[aps,prl,twocolumn,showpacs,amsmath,amssymb]{revtex4}
\usepackage{graphicx}% Include figure files
\usepackage{dcolumn}% Align table columns on decimal point
\usepackage{bm}% bold math

\begin{document}

\title{Exciton-phonon effects in carbon nanotube optical absorption}

\author{Vasili Perebeinos, J. Tersoff, and Phaedon Avouris$^*$}
\affiliation{IBM Research Division, T. J. Watson Research Center,
Yorktown Heights, New York 10598}

\date{\today}

\begin{abstract}
We find that the optical properties of carbon nanotubes reflect
remarkably strong effects of exciton-phonon coupling.
Tight-binding calculations show that a significant fraction of the spectral
weight of the absorption peak is transferred to a distinct exciton+phonon
sideband, which is peaked at around 200 meV above the main
absorption peak.
This sideband provides a distinctive signature of the excitonic
character of the optical transition.
The exciton-phonon coupling is reflected in a
dynamical structural distortion, which contributes a binding
energy of up to 100 meV. The distortion is surprisingly
long-ranged, and is strongly dependent on chirality.
\end{abstract}

\maketitle

The optical properties of carbon nanotubes are currently the focus
of intense experimental and theoretical attention
\cite{Li,Connell,Bachilo,Hagen,Lebedkin,Lefebvre,Ando,%
Pedersen,Kane,Louie,Chang,Perebeinos}, and even single nanotube
electro-optical devices have been demonstrated
\cite{Misewich,Marcus,Marcus2}.
Most of the experimental results have been
discussed and analysed in terms of interband transitions.
In particular, nanotube bandgaps were determined in this way.
However, theoretical
calculations of the optical spectra
\cite{Louie,Perebeinos,Chang} suggest that the
observed transitions correspond to exciton energies,
not interband transitions.  Moreover, the exciton
binding energies are anomalously large in nanotubes
\cite{Ando,Pedersen,Louie,Chang,Perebeinos},
corresponding to a substantial fraction of the bandgap,
so the optical transition energy is quite different
than the bandgap.

If correct, these theoretical results require a reevaluation of
our current picture of nanotube electronic structure. Device
properties are particularly sensitive to the bandgap, because of
the central role of tunneling in most nanotube transistors
\cite{HeinzeSB}. Thus a definitive test of the excitonic
interpretation of optical transitions is needed. Fine structure in
the optical data can provide a distinctive signature, facilitating
comparison between different models to unambiguously verify the
role of excitonic transitions. Recent photoconductivity excitation
spectra \cite{Avouris,Freitag} show sidebands at about 200 meV
above to the main absorption peaks, suggesting the involvement of
phonons. Fluorescence excitation spectroscopy also shows structure
at the phonon energy, allowing more detailed analysis with finer
energy resolution \cite{SGChou}. But while the interaction of
single electrons with phonons has been studied extensively, we
know of no studies of exciton-phonon interactions in nanotubes.

\begin{figure}
\includegraphics[height=6.23in,width=2.79in]{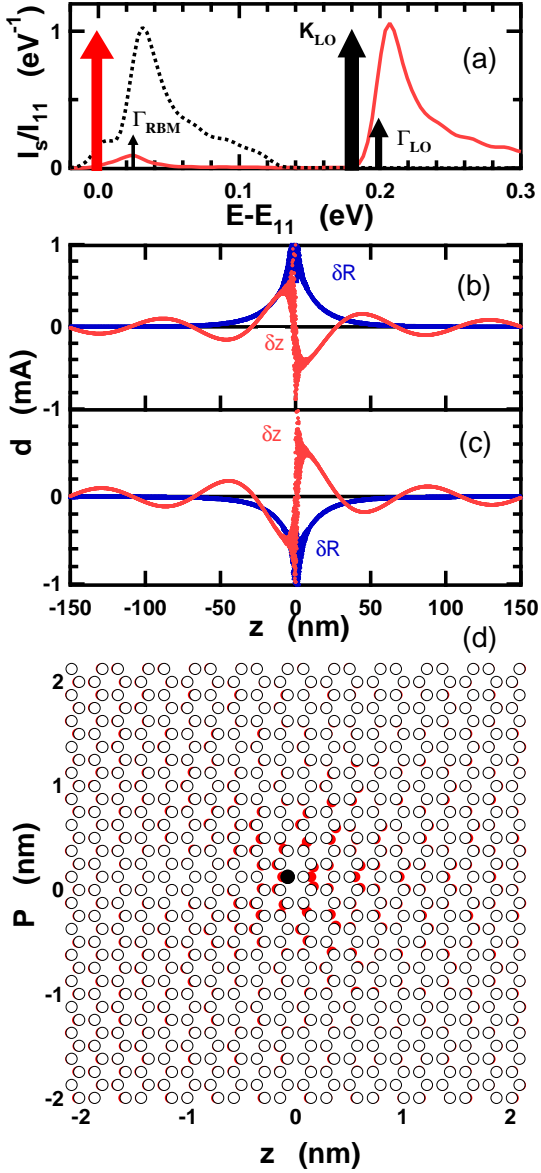}
\caption{\label{fig1} (color online). (a) Absorption spectrum
Eq.~(\protect\ref{eq4}) for (17,0) tube and $\varepsilon=2$ (solid
red curve and red delta-function). Dotted line shows electronic
contribution to sideband energy (with 10 meV Gaussian broadening).
Phonon contributions are quite narrow and are shown schematically
as delta-functions (height $\sim I^{1/2}$). (b) Long-range tail of
the atomic displacements with respect to the hole position (at
origin) for a (16,0) tube: radial displacements (labelled $\delta
R$, in blue), and displacements parallel to the axis (labelled
$\delta z$, in red). (c) Same for (17,0) tube. (d) Short-range
azimuthal and axial distortions, shown on ``unwrapped'' carbon
nanotube by displaying the displaced atom positions (red circles)
together with undisplaced positions (white circles), with
displacements magnified by a factor of 230 for visibility. Hole
position is solid circle at center.}
\end{figure}

\begin{figure}
\includegraphics[height=2.77in,width=3.28in]{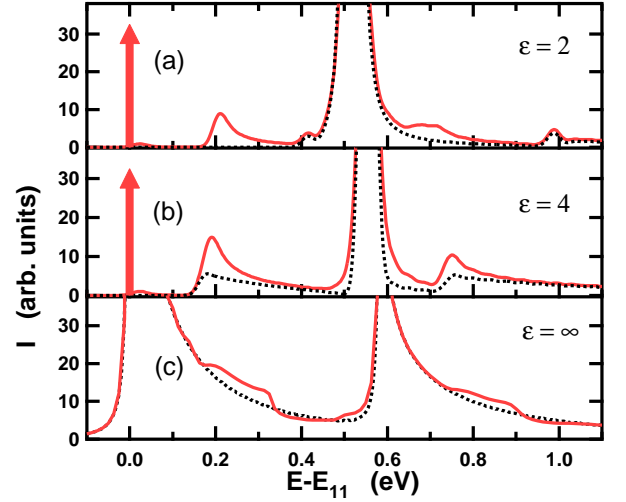}
\caption{\label{fig2} (color online). Absorption spectra in (17,0)
tube calculated with and without electron-phonon couplings (solid
red and dashed black curves respectively) for (a) $\varepsilon=2$,
(b) $\varepsilon=4$, (c) $\varepsilon \rightarrow \infty$. The
zero of energy here corresponds to the onset of the first
optically active exciton, which has zero width and is shown by the
vertical arrows. The onset of continuum (bandgap) transitions is
at 0.48 eV, 0.19 eV, and 0 in (a), (b), and (c) respectively. The
width of the second exciton is finite due to: (1) coupling with
the first band electron-hole continuum to give lifetimes of
$\tau_{\rm ee}$=10 fs and 15 fs for $\varepsilon=$2 and 4
respectively; and (2) coupling with phonons to give $\tau_{\rm
ph}$=90 fs and 33 fs for $\varepsilon=$2 and 4 respectively. Delta
functions in Eq.~(\protect{\ref{eq5}}) were Gaussian broadened
with width of 20 meV, except in (c) we use 6.5 meV to emphasize
that the width of the continuum peak is intrinsic.}
\end{figure}

We therefore investigate theoretically the role of electron-phonon
coupling in the optical spectra of nanotubes, including a
comparison of between excitons and free-carrier transitions. We
find surprisingly strong phonon effects in the excitonic spectra.
Dynamical effects lead to the transfer of a significant fraction
of the spectral weight from the exciton absorption peak (zero
phonon line) to a phonon sideband peaked at around 200 meV
(Fig.~\ref{fig1}a and Fig.~\ref{fig2}a-b), consistent with the
experimental observations.  In contrast, our calculations for
electron-phonon interaction without excitonic binding
(Fig.~\ref{fig2}c) do not show a distinct phonon sideband. These
results clearly suggest that the optical data must be interpreted
in terms of excitonic transitions, and do not provide a direct
measure of the bandgap.

We find that the fraction of the intensity transferred to the
phonon sideband is inversely proportional to the nanotube
diameter. The exciton-phonon binding energy is also unexpectedly
high, about 60-100 meV. The associated lattice distortions exhibit
an intriguing structure, they extend far beyond the exciton itself
\cite{Perebeinos} and reverse sign for different nanotube
chiralities (Fig.~\ref{fig1}b-c). The effect of the
dielectric environment on the spectra is also discussed.  (We
treat individual nanotubes, as in fluorescence excitation
spectroscopy. Nanotube bundles are less suitable for measuring
detailed lineshapes.)

In the absence of exciton-phonon coupling, emission or absorption
of a photon involves an exciton of total wavevector $q_{\phi}$
corresponding to the photon momentum, hereafter approximated as
$q_{\phi}=0$. The exciton-phonon coupling mixes this exciton with
phonons and with excitons of other $q$, such that the total
exciton+phonon momentum is conserved.
The finite-$q$ exciton wavefunction can be found from the solution
of the Bethe-Salpeter Equation (BSE) \cite{Rohlfing} in the basis
of a tight-binding Hamiltonian \cite{Saito}, analogous to the
$q=0$ case \cite{Perebeinos}:
\begin{eqnarray}
\left\vert\Psi_{q}^S\right\rangle=\sum_{k}
A^S_{kq}u^{\dagger}_{k+q}v_{k}\left\vert{\rm GS}\right\rangle.
\label{eq1}
\end{eqnarray}
Here $A^S_{kq}$ is the eigenvector of the $S$'s state of BSE
solution; $u^{\dagger}_{k+q}$ ($v_{k}$) creation (annihilation) of
an electron in the conduction (valence) band acting on the ground
state $\left\vert{\rm GS}\right\rangle=\prod_k
v^{\dagger}_k\left\vert{\rm vac}\right\rangle$. The indices $k$ and $q$
each label both the continuous 1D wavevector along the tube axis and
the discrete circumferential wavevector.

We model the electron-phonon interaction by the
Su-Schrieffer-Heeger (SSH) model \cite{Su}, with matrix element
$t=t_0-g\delta u$ dependent on the change of the nearest neighbor
C-C distance ($\delta u$), where $t_0=3$ eV. We take the
electron-phonon coupling constant to be $g=5.3$ eV/\AA~ as
predicted theoretically for a related molecular problem
\cite{Perebeinos2}, consistent with fits to the Peierls gap in
conjugated polymers \cite{Peierls}.

After Fourier transformation the intraband SSH Hamiltonian has the
form:
\begin{eqnarray}
{\cal H}_{\rm el-ph}=\sum_{kq\mu}{\rm
M}_{kq}^{\mu}(v^{\dagger}_{k+q}v_{k}-u^{\dagger}_{k+q}u_{k})
(a_{q\mu}+a_{-q\mu}^{\dagger}),
\label{eq2}
\end{eqnarray}
where ${\rm M}_{kq}^{\mu}\propto g N^{-1/2}$ is momentum dependent
electron-phonon coupling; $a^{\dagger}_{-q\mu}$ is a phonon
creation operator with wavevector $-q$ and phonon band index
$\mu=1...6$; and $N$ is the number of primitive unit cells, each
containing two carbons.
For the phonon spectrum we used
a force-constant model similar to Saito {\it et al.}
\cite{Saito2}.

The electron-phonon Hamiltonian mixes the optically active
$q$=0 exciton with finite-$q$ excitons in combination with
phonons of wavevector $-q$:
\begin{eqnarray}
{\cal H}_{\rm
el-ph}\left\vert\Psi_0^S\right\rangle&=&-\sum_{S'q\mu}
B^{SS'}_{q\mu}
a_{-q\mu}^{\dagger}\left\vert\Psi_q^{S'}\right\rangle \nonumber \\
B^{SS'}_{q\mu}&=&\sum_k {\rm
M}^{\mu}_{kq}A_{kq}^{S'*}(A_{k,0}^{S}+A_{k+q,0}^{S}) \label{eq3}
\end{eqnarray}
Here the orthogonality relation of the BSE solution $(\sum_S
A_{kq}^SA_{k'q}^{S*}=\delta_{kk'})$ has been used to derive
exciton-phonon coupling amplitudes $B^{SS'}_{q\mu}$.

The wavefunction and the spectral line shape can be evaluated in
second order perturbation theory for the lowest optically active
exciton $s$:
\begin{eqnarray}
\left\vert\tilde{\Psi}_0^s\right\rangle&\propto&
\left\vert\Psi_0^s\right\rangle+\sum_{qS'\mu}\frac{B_{q\mu}^{sS'}}{E_{q}^{S'}+
\hbar\omega_{-q\mu}-E_0^s}a_{-q\mu}^{\dagger}\left\vert\Psi_{q}^{S'}\right\rangle
\nonumber \\
I(\omega)&\propto&\delta(E_0^{s}-\hbar\omega)+
\sum_{qS'\mu}\frac{\left\vert
B^{sS'}_{q\mu}\right\vert^2}{(E_q^{S'}+\hbar\omega_{-q\mu}-E_0^{s})^2}
\nonumber \\
&&\times\delta(E_q^{S'}+\hbar\omega_{-q\mu}-\hbar\omega)
\label{eq4}
\end{eqnarray}

The calculated absorption spectrum Eq.~(\ref{eq4}) is shown in
Fig.~\ref{fig1}a (solid red curve) for a (17,0) tube and
$\varepsilon=2$, where $\varepsilon$ is the dielectric constant of
the embedding medium \cite{Perebeinos}. Because of exciton-phonon
coupling, the main absorption peak looses 8\% of it's spectral
weight to the sideband, which corresponds to the continuum of
finite-$q$ excitons plus phonon of wavevector $-q$. Most of the
transferred spectral weight goes to the prominent sideband at
about 210 meV above the zero phonon line, with 4\% of the spectral
weight falling between 100 and 300 meV. The SSH Hamiltonian has
little coupling to the low frequency modes and we find a much
weaker replica at the radial breathing mode (RBM) frequency
[i.e. 20 meV for a (17,0) tube].

To understand the spectrum in more detail, we decompose
the sideband energy into contributions from phonon energy
and exciton dispersion.  Specifically, for each transition
within 300 meV above the zero-phonon line,
we project out the energy contribution from phonons
{\it vs}  electronic excitation (due to the admixture of
finite-$q$ excitons).
The phonon contribution to the sideband exhibits three peaks
Fig.~\ref{fig1}a, corresponding to the longitudinal optical (LO) phonon
band edges at the $K$ and $\Gamma$ points of the graphene
Brillouin zone, and the radial breathing mode. The $K$ phonon
dominates.  It has stronger coupling; and more importantly, it
mixes exciton bands \cite{Perebeinos3}, allowing absorption by the
``dark'' (dipole forbidden) bands \cite{Perebeinos,Zhao}.
The energy difference
between the sideband peak position and optical phonon frequency is
due to the exciton ``recoil energy'', i.e.\ the energy from finite
exciton $q$, which contributes $\sim$30 meV here.

In the presence of an exciton the nanotube distorts dynamically. While the
electron remains within roughly 2 nm of the hole for the cases
shown in Fig.~\ref{fig1}, the structural distortions are far more
long-ranged. The distortions calculated using the wavefunction
Eq.~(\ref{eq1}) are shown in Fig.~\ref{fig1}b and \ref{fig1}c for
(16,0) and (17,0) tubes respectively, plotting the atomic
displacements relative to the position of the hole.
The breathing distortions decay exponentially away from the
exciton, with a decay length of $\lambda_{b}=13$ nm. We find that
$\lambda_b$ is proportional to the tube diameter. The sign of the
breathing distortion depends on chirality indices (n,m): positive
for mod(n-m,3)=1 and negative for mod(n-m,3)=2. From the sign of
the exciton-phonon matrix elements, we expect a reversal of the
signs of the breathing distortions for the second exciton. The
displacements parallel to the tube axis decay far more slowly even
than this, oscillating with a wavevector $\lambda_z\approx 80$ nm,
which appears to be insensitive to the tube diameter.
The short-range distortions near the exciton are shown in
Fig.~\ref{fig1}d.

The spectra for higher-energy excitons
($E_0^{S}>E_0^{s}$) cannot be obtained from Eq.~(\ref{eq4}),
because the denominator $E_q^{S'}+\hbar\omega_{-q\mu}-E_0^{S}$ can
be arbitrarily close to zero and the perturbation theory breaks
down. Toyozawa showed \cite{Toyozawa} that an exact solution for
the absorption spectra
has a form similar to the perturbation theory expression, with an
energy dependent lifetime and polaronic shift in the energy
denominator. We approximate Toyozawa's solution by evaluating the
lifetime broadening in the Random Phase Approximation (RPA):
\begin{eqnarray}
I(\omega)&=&\sum_{S}\frac{
f_S}{\pi}\frac{\Gamma_S(\omega)}{(\hbar\omega-E_0^{S})^2+\Gamma_S(\omega)^2}
\nonumber \\
\Gamma_S(\omega)&=&\pi\sum_{S'q\mu}\left\vert
B^{SS'}_{q\mu}\right\vert^2\delta(\hbar\omega-E_q^{S'}-\hbar\omega_{-q\mu}),
\label{eq5}
\end{eqnarray}
where $f_S$ is the oscillator strength of the $S$ exciton
\cite{Perebeinos}. We checked that the Toyozawa solution obtained
with a self consistent Born approximation neglecting the $k$
dependence of the self energy
$\Sigma_S(E)=\Delta_S(E)+\Gamma_S(E)$ does not change the RPA
result Eq.~(\ref{eq5}). If $\hbar\omega=E_0^S$ in Eq. (\ref{eq5})
then $\Gamma_S(E_0^S)/\hbar$ equals half of the reciprocal
lifetime of $S$-exciton due to the scattering by phonons. The
binding energy shift for the $S$-exciton in RPA approximation is:
\begin{eqnarray}
\delta E_S=\sum_{S'q\mu}\left\vert
B^{SS'}_{q\mu}\right\vert^2{\cal
P}(E_0^S-E_q^{S'}-\hbar\omega_{-q\mu})^{-1}, \label{eq6}
\end{eqnarray}
where ${\cal P}$ denotes the principle part.

The absorption spectra calculated with and without exciton-phonon
coupling are shown in Fig.~\ref{fig2} for the tube embedded in
dielectric $\varepsilon=2$, 4, and for the free electron-hole pair
absorption (equivalent to the limit $\varepsilon \rightarrow
\infty$). In the absence of exciton-phonon interactions there are
two strong exciton absorption peaks, each followed by the
corresponding continuum of intraband absorption \cite{Perebeinos}.
The first exciton has zero width in this approximation, and is
shown by a vertical arrow in Fig.~\ref{fig2}a and \ref{fig2}b.
Both exciton lines show a distinct phonon sideband about 200 meV
above the main absorption line.
(For $\varepsilon=4$
the exciton binding energy is about 0.2 eV, so the band-to-band
absorption (dotted line) also contributes to the sideband
intensity in Fig. \ref{fig2}b.)
In contrast, there is no distinct
phonon peak associated with band-to-band absorption
(Fig.~\ref{fig2}c). Thus the phonon peak provides a clear
signature of whether or not the absorption is excitonic in nature.

The second exciton can decay
into free electron-hole pairs of the first band via Coulomb
interaction or by emitting a phonon, giving rise to a finite
lifetime of the second exciton resonance.
The electronic and phonon contributions are given in the figure
caption for a (17,0) tube for two values of $\varepsilon$.
The actual value of the lifetime,
as well as other details of the spectrum,
are very sensitive to the lineup of the resonance peak
position with the onset of the first band continuum,
and thus are also sensitive to the
tube radius and dielectric environment.
Therefore experimental measurements of the first exciton
are easier to compare with theory.

\begin{figure}
\includegraphics[height=2.07in,width=3.63in]{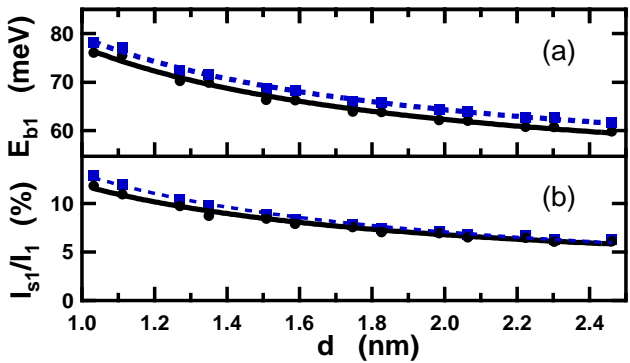}
\caption{\label{fig4}  (color online). (a) The phonon binding
energy, and (b) fraction of the spectral weight transferred to the
exciton+phonon sideband, for the first exciton in zig-zag tubes,
versus tube diameter for $\varepsilon=2$ (black circles) and for
$\varepsilon=4$ (blue squares) respectively. The curves (black
solid for $\varepsilon=2$ and blue dashed for $\varepsilon=4$) are
fits $A_{b,I}+B_{b,I}/d$, where $A_b=47$ (meV), $B_b=30$ (meVnm)
and $A_I=1.7$ (\%), $B_I=10$ (\%nm) for $\varepsilon=2$.}
\end{figure}

The total exciton-phonon binding energy calculated from Eq.~(\ref{eq6})
has the largest contribution from the coupling to the higher
energy states, which have smaller optical spectral weight due to the
energy denominator in
Eq.~(\ref{eq5}). The dependence of the binding
energy on the diameter $d$ is shown in Fig.~\ref{fig4}a,
along with a phenomenological fit,
for the first exciton in different zig-zag tubes,
for  $\varepsilon=2$ and $4$.
The second exciton binding energy is
always larger than the first exciton, by 15-30\%, due to the larger
effective mass of the former. The fraction of the spectral weight
transfer for the first exciton Eq.~(\ref{eq4}) is shown on
Fig.~\ref{fig4}b, along with a fit for the same values of
$\varepsilon$.

In conclusion, we calculate exciton-phonon effects
in the absorption spectra of carbon nanotubes,
predicting a significant spectral weight transfer from the main
(excitonic) absorption peak to a phonon sideband at around 200 meV
above the zero phonon line, for both the first and second excitons.
Comparable structure is seen in recent experiments
\cite{Avouris,Freitag,SGChou}.
In contrast, for band-to-band transitions we find that there is
no distinct sideband;  instead the phonon structure represents
only a slight perturbation of the spectrum.
We therefore believe that the phonon sideband
provides direct experimental evidence, that the optical transition
is excitonic in nature and cannot be used as a
direct measure of the bandgap.

\end{document}